\documentclass[prx,aps,english,twocolumn,notitlepage]{revtex4-1} 
\usepackage[T1]{fontenc} 
\usepackage{sidecap}
\usepackage[latin1]{inputenc}
 \usepackage{babel} \usepackage{txfonts}
 \usepackage{epsfig,epsf,psfrag} 
\usepackage{graphicx} 
\usepackage{tcolorbox}
\usepackage{framed}
\usepackage{booktabs,xcolor}
\usepackage{xcolor,colortbl}
\usepackage{afterpage}

\definecolor{Gray}{gray}{0.85}
\definecolor{LightCyan}{rgb}{0.88,1,1}
\definecolor{darkpink}{rgb}{0.91, 0.33, 0.5}
\newcolumntype{a}{>{\columncolor{Gray}}c}
\newcolumntype{b}{>{\columncolor{white}}c}
\usepackage{subfigure}
\usepackage{pslatex}
\usepackage{epic,eepic} 
\usepackage{color,pstcol}
\usepackage{pstricks} 
\usepackage{fancyhdr}  \parindent0em  
\def\ua{\uparrow}
\def\da{\downarrow}

\begin{document} 
  \title{A scheme to realize the quantum spin-valley Hall effect in monolayer graphene} 
  \author{SK Firoz Islam}
 \author{Colin Benjamin} \email{{\bf Corresponding Author. Email: colin.nano@gmail.com} }\affiliation{ National institute of Science education \& Research, Bhubaneswar 751005, India }

\begin{abstract}
Quantum spin Hall effect was first predicted in graphene. However, the weak spin orbit interaction
in graphene meant that the search for quantum spin Hall effect in graphene never fructified.
In this work we show how to generate the quantum spin-valley Hall effect in graphene via
quantum pumping by adiabatically modulating a magnetic impurity  and an electrostatic potential in
a monolayer of strained graphene. We see that not only exclusive spin polarized  currents can be pumped in the two valleys in  exactly opposite directions but one can have pure spin currents flowing in opposite
directions in the two valleys, we call this novel phenomena the quantum spin-valley Hall effect.  This means that the twin effects of quantum valley Hall and quantum spin Hall can both be probed simultaneously in the proposed device. This work will significantly advance the field of graphene spintronics, hitherto hobbled by the lack of spin-orbit interaction. We obviate the need for any spin orbit interaction and show how graphene can be manipulated to posses features exclusive to topological insulators.
\end{abstract}

\maketitle
\section{Introduction}
Graphene is the material of the 21st century, what Silicon was to the 80's and 90's. It continues to be the most exciting
material in condensed matter today, although challenged by topological insulators, for it's ability to show some striking
unusual phenomena and it's potential applications in nanoelectronics\cite{neto}. Several remarkable features of graphene,
which are in complete contrast to semicon ductor heterostructures, are Klein tunneling\cite{novoselov} and room temperature
quantum Hall effect\cite{geim}. It's electronic properties are governed by massless linear dispersion- Dirac behavior at low
energy around two distinct valleys $K$ and $K'$ in it's Brillouin zone. These two valleys, connected by time reversal symmetry, can also act as
an additional degree of freedom just like spin in spintronics\cite{moodera}. Similar to spintronics, the
valley degree of freedom can also be exploited as regards applications in quantum computation- referred as
valleytronics\cite{valley1,valley2,valley3}. In valleytronics proposals, via controlling the valley degree of
freedom, valley based filter, valve and field effect transistor have been already reported\cite{valley3,fet,VHE,vdf,vdf1,vdf2}.
There were also proposals of quantum spin valley Hall effect in multilayer graphene\cite{macdonald}, spin-valley filter in graphene \cite{peeters_svf} and thermally driven spin and valley currents in Group-VI dichalcogenides\cite{jauho}.

An exciting aspect of graphene is that a mechanical strain provides an excellent way to control valley degree of freedom. Strain causes
an opposite transverse velocity in the two valleys (K,K')\cite{bao,katsnelson}.
The separation in momentum space between two valleys, generated by the opposite velocity, 
causes the well known valley Hall effect\cite{tony1,tony2}. The various Hall effects possible in graphene are mentioned in the Box.
Apart from strain, there are several other proposed schemes to produce valley
polarization-like triangular wrapping effects\cite{valley3}, edge effects in graphene nanoribbons\cite{valley4} and a
valley dependent gap generated by substrate\cite{mass1,mass2,mass3}, etc. Strained graphene can also show some electro-optic properties like:
total internal reflection, valley dependent Brewster angle and Goos Hanchen effects\cite{peeters}.

\fbox {\begin{minipage}{26em}\textbf{\hspace{1em}The possible Hall effects in monolayer graphene}\\
 Depending on the situation encountered one can have any or some of the following conditions satisfied in our proposed device:\\
Ia. $I_c^{K}=I_c^{K'}=0$ ($I_{\ua}=-I_{\da}$)- The condition of pure spin current generation in each valley
regardless of the angle of incidence of electron. Here, $I_{c}^{K/K'}=I^{\uparrow}_{K/K'}+I^{\downarrow}_{K/K'}$, the total charge current in $K/K'$
valley,\\
Ib. $I_c^{K}(\phi)=I_c^{K'}(\phi)=0$- The condition of pure spin current generation in each valley
at a particular angle of incidence $\phi$.\\
II. $I_{c}^{K}(\phi)=-I_{c}^{K'}(\phi)$, charge currents are same and opposite in each valley for a particular angle of
incidence-quantum valley Hall effect (QVH).\\
IIIa. $I^{K}_{\ua}(\phi)=-I^{K'}_{\da}(\phi)$ with $I^{K}_{\da}(\phi)=I^{K'}_{\ua}(\phi)=0$ i.e; two valleys carrying opposite spin current
with same magnitude but in opposite direction-quantum spin-valley Hall effect (QSVH) of 1st kind,\\
IIIb. $I^{K}_{\da}(\phi)=-I^{K'}_{\ua}(\phi)$ with $I^{K}_{\ua}(\phi)=I^{K'}_{\da}(\phi)=0$-QSVH of 1st kind.\\
IV. $I^{K}_{\ua}(\phi)-I^{K}_{\da}(\phi)=-[I^{K'}_{\ua}(\phi)-I^{K'}_{\da}(\phi)]$,
QVH with pure spin current in each valley. This can also be termed as QSVH of 2nd kind.
\end{minipage}}

In the present work, we use the following symbols for different components of pumped currents: spin-up current: $I_\uparrow$ , spindown
current: $I_\downarrow$, spin current: $I_{s} = I_{\uparrow} - I_\downarrow$ and charge current: $I_{c} = I_{\uparrow} + I_{\downarrow}$. Quantum spin-valley Hall effect (QSVH) is defined as one
valley carries a current of only spin up (spin down) and the
other valley carries a current of spin down (spin up) with same
magnitude but in exactly opposite direction. A variant of this,
i.e., two valleys carry pure spin currents in exactly opposite
direction with same magnitude is termed as QSVH of 2nd
kind, see Fig.1 for a pictorial on QVH and QSVH (first and
second kinds).
In this work, we aim to manipulate both degrees of freedom, i.e.,
spin and valley, for which we dope the graphene monolayer
with a magnetic impurity and an electrostatic potential and
also apply an in-plane strain to the graphene layer.
We find that the condition (Ia) of pure spin current generation
in each valley is satisfied in Fig.~7. The condition (Ib) of pure
spin current generation at a particular angle of incidence and the QVH appear in Fig.~5(b) (upper panel). We get the condition
(III) of QSVH of 1st kind in Fig.~4, while the condition (IV)
of QSVH of 2nd kind is found in Fig.~5(b) (lower panel).

\begin{figure}
    \includegraphics[width=0.48\textwidth,height=60 mm]{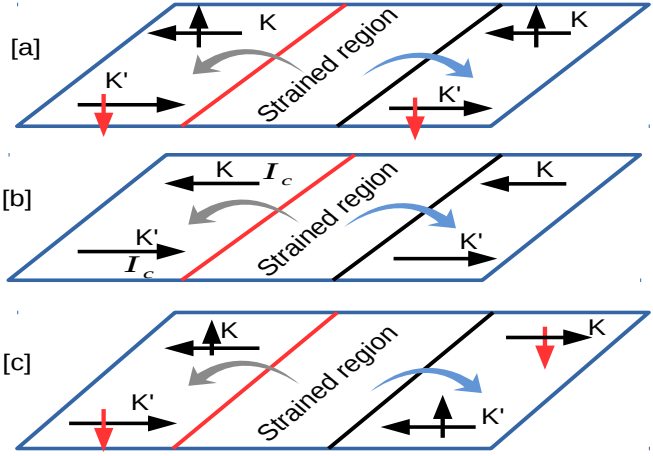} 
    \caption{A pictorial representation of different possible charge/spin pumped currents in each valley.
The red and black solid lines are two scatterers (magnetic impurity and delta potential). In Fig.~1(a), K and
K' valley carry exclusively spin up and spin down currents in exactly opposite direction-quantum spin valley Hall effect (QSVH of 1st kind).
Fig. (1b) shows charge currents in each valley are same in magnitude and opposite in direction-quantum valley Hall effect (QVH).
Fig. (1c) shows pure spin currents in each valley are same and opposite in direction, we call it quantum valley Hall effect with pure
spin current (QSVH of 2nd kind).
          } \label{fig-qsvh}
\end{figure}

\begin{figure}
\includegraphics[width=0.55\textwidth,height=60 mm]{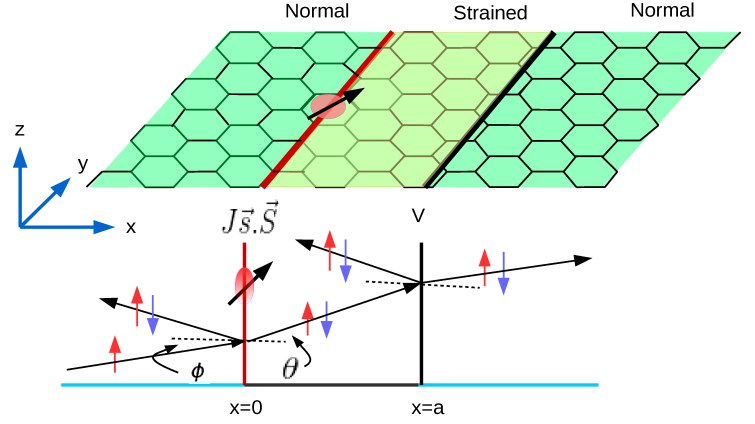}
\caption{Top: The graphene layer with the red solid line representing the magnetic impurity at $x=0$, while black line is for electrostatic delta potential at $x=a$.
The interveneing portion is the strained region. Valley and spin dependent currents are pumped out of the strained region by modulating
magnetic impurity and electrostatic potential. The lower picture shows incident up electron (for K-valley) is
reflected/transmitted with or without spin flip by magnetic impurity. The angle of incidence is $\phi$, while the angle of refraction into the
strained region is $\theta$ for a particular valley. Similar phenomena occurs at the other interface with electrostatic delta potential
without spin flip.}
\label{Fig1}
\end{figure}

\section{Theory}
Graphene is a two dimensional carbon allotrope with hexagonal lattice structure~\cite{neto} that can be split into two
triangular sublattices $A$ and $B$.  
    We consider a mechanical strain to be applied to the graphene sheet which is lying in the $x$-$y$ plane\cite{bao,katsnelson},
in the region between magnetic impurity at $x=0$ and electrostatic potential at $x=a$. The 
sketch of the considered system is shown in Fig. \ref{Fig1}. 
Strain is included  in the Dirac Hamiltonian as follows- In-plane mechanical strain affects the 
hopping amplitude between two nearest neighbors and can be described as a gauge vector which are
opposite in two valleys.
In the Landau gauge, the vector potential corresponding to the strain is ${\bf A}=(0,A_y)$.
The system can be easily described by the Hamiltonian\cite{castro,peeters,njp}, as:
\begin{equation}\label{hamil}
 \mathcal{H}_{K/K'}=H_{K/K'}+J{\bf s}.{\bf S}\delta(x)+V\delta(x-a)
\end{equation}
with $H_{K}=\hbar v_F{\bf \sigma.(k-t)}$ and $H_{K'}=\hbar v_F{\bf \sigma^{\ast}.(k+t)}$.
Here, $ t= \frac{A_y}{\hbar v_F} [\Theta(x)-\Theta(x-a)]$ is the strain with $\Theta$ being the step function,
$v_F$ is the Fermi velocity. The first term represents the kinetic energy for graphene with ${\bf \sigma}=(\sigma_x,\sigma_y)$-
the Pauli matrices  that operate on the sublattices $A$ or $B$ and ${\bf k}=(k_x,k_y)$ the 2D wave vector.
Second term is the exchange interaction between Dirac electron and magnetic impurity and final term
is an electrostatic delta potential.  In the second term $J$ represents the strength of 
the exchange interaction which depends on the magnetization of the magnetic impurity and modulating it's magnetization one 
can effectively change $J$. The spin of Dirac electron is denoted by $s$, while $S$ represents spin of the magnetic impurity.
V is the strength of the potential, situated at $x=a$. Energy of the electron will be denoted by 'E'.

A short review of basic theory of quantum pumping and the method of solving the 
scattering problem for spin-up/down electron is given in the following sub-sections.
\subsection{Quantum pumped currents}
 Adiabatic quantum pumping is a phenomena in which charge can be transported
without any external bias. It requires cyclic variations of the scattering matrix,
which could be realized by the periodic modulation of two independent system parameters of the device. The first
experimental attempt at quantum pumping was done by M. Switkes in 1999\cite{switkes}, where 
the pumping signal was recorded in response to the cyclic deformation of the confining
potential. Concurrently, P.W. Brouwer provided the theory of quantum pumping\cite{brouwer}. 
Pumping has also been used to generate spin 
dependent currents in theory\cite{benjamin} as well as experiments\cite{watson}. Recently, several theoretical works
have been reported on quantum spin and valley current pumping in graphene based devices consisting of
ferromagnet/valley dependent mass term and gate electrodes\cite{lin,jin}. To generate pure spin current,
spin-up and spin-down currents have to be exactly same in magnitude and opposite in direction.
Same goes for pure valley currents also. To calculate quantum pumped currents, we proceed as follows:
The infinitesimal change of two system parameters, say $\zeta_i$ with $i=1,2$,
causes an infinitesimal charge transport ($dQ$) through the lead $\alpha$(say)- in a particular valley($K$)  with spin ($\tau$) is given by-
\begin{eqnarray}
dQ^{\tau \alpha}_{K}(t)=e \sum_{i}\frac{dN_{\tau \alpha}}{d\zeta_i} \delta \zeta_{i}(t)
\end{eqnarray}

and the current transported in one period being-

\begin{eqnarray}
I^{\tau \alpha}_{k}=\frac{ew}{2\pi}\int_{0}^{2\pi/w} dt \sum_{i}\frac{dN_{\tau \alpha}}{d\zeta_i}\frac{d\zeta_{i}}{dt}\label{eq:Ipump},
\end{eqnarray}
$w$ being the frequency of applied modulation to parameters $\zeta_i$.
The quantity $dN_{\tau\alpha}/d\zeta_i$ is known as emissivity which can be obtained from
the elements of the scattering matrix, in the zero temperature limit
by-

\begin{eqnarray}
\frac{dN_{\tau \alpha}}{d\zeta_i}=\frac{1}{2\pi}\sum_{\tau^{\prime} , \beta} \Im (\frac{\partial s^{\tau \tau^{\prime}}_{\alpha \beta}}{\partial \zeta_i}
s^{\tau \tau^{\prime} *}_{ \alpha \beta}).
\label{eq:emis}
\end{eqnarray}

Here $s^{\tau\tau^{\prime}}_{ \alpha \beta}$ represents the scattering
matrix elements as denoted above, $ \alpha$ and $\beta$ 
take values $1$ (for pumping to left of strained region) and $2$ (for pumping to right of strained region), while $\tau, \tau^{\prime}$ are the spin indices, $\ua$ and $\da$, depending on
whether spin is up or down. The symbol ``$\Im$'' indicates the
imaginary part of the complex quantity inside parenthesis. $s^{\tau\tau^{\prime}}$  indicates scattering
amplitudes when incident electron with spin index $\tau^{\prime}$ is scattered (reflected or transmitted) to
the state in spin index $\tau$.

The individual spin pumped currents are generated by adiabatically modulating the magnetization of impurity 'J'
and the strength of the electrostatic ``delta'' potential $V$, herein
$\zeta_{1}=J=J_{0}+J_{p}\sin(wt)$ and
$\zeta_{2}=V=V_{0}+V_{p}\sin(wt+\Omega)$.  $w$ as before is the frequency of modulation
and $\Omega$ is the phase difference between the two modulated parameters. 
A section on the feasibility of experimental realization of the proposed device is given in the conclusion.

The line integral of Eq. (\ref{eq:Ipump}) can be converted into an surface integral by
using Stokes theorem on two dimensional plane. Then after some straight forward manipulation,
%
%
for sufficiently weak pumping ($\delta \zeta_{i} \ll \zeta_{i}$), we have (see for details \cite{benjamin}),

\begin{eqnarray}\label{stokes1}
I^{\tau \alpha}_{K}(\phi)=\frac{ew\delta \zeta_{1}\delta \zeta_{2}\sin(\Omega)}{2\pi} \sum_{\beta=1,2} 
\Im (\frac{\partial s^{\tau \tau^{\prime}*}_{ \alpha \beta}}{\partial \zeta_1} \frac{\partial s^{\tau \tau^{\prime}}_{ \alpha \beta}}{\partial \zeta_2}).
\end{eqnarray}
Weak pumping is defined by: $J_{p} \ll J_{0}, V_{p} \ll V_{0}$, and Eq.~\ref{stokes1} reduces to-

 \begin{equation}\label{pump}
I^{\tau \alpha}_{K}(\phi)=I_{0} \sum_{\tau^{\prime}=\ua,\da , \beta=1,2}
\Im (\frac{\partial s^{\tau\tau^{\prime} *}_{\alpha \beta}}{\partial J} \frac{\partial s^{\tau \tau^{\prime}}_{\alpha\beta}}{\partial V}), \mbox {wherein  }
I_{0}=\frac{ew J_{p}V_{p} \sin(\Omega)}{2\pi}.
\end{equation}

For parameter values $e=1.6 X 10^{-19}$ Coulombs, $w=10^{8}$ Hertz from Ref.\cite{switkes}, $I_0$ is of the order of $J_{p}V_{p} 10^{-11}$ Amperes, with $J_{p}$ and $V_{p}$ again defined as above but in their dimensionless form. Since we are in the weak pumping regime we can consider $J_p$ and $V_p$ to be each around $0.1$ as in Figs.~5-9 we have taken $J=2V=2 eV-nm$, this makes $I_{0} =  10^{-13}$ Amperes. We are considering  pumped spin currents into lead 1 (left of strained region), therefore $\alpha=1$ throughout this paper.
%
In the above equation, if we consider pumped currents in $K$ valley to left of strained region then $alpha=1$ with spin $\tau=\uparrow$ then  different scattering amplitudes are denoted by-

$s^{\ua\ua}_{11}\equiv r_{\ua\ua}$, $s^{\ua\da}_{11}\equiv r_{\ua\da}$, $s^{\ua\ua}_{12}\equiv t^{\prime}_{\ua\ua}$,
and $s^{\ua\da}_{12}\equiv t^{\prime}_{\ua\da}$, where\\
 $r_{\ua\ua}$: reflection amplitude for spin-up electron reflected to the spin-up state,\\
 $r_{\ua\da}$: reflection amplitude for spin-down electron reflected to the spin-up state,\\
 $t^{\prime}_{\ua\ua}$: transmission amplitude for spin-up electron transmitted to the spin-up state, and\\
 $t^{\prime}_{\ua\da}$: transmission amplitude for spin-down electron transmitted to the spin-up state.\\
 Similarly, we can calculate the spin down current by replacing $\ua\rightarrow\da$ and vice-versa.
 
%
Here, $s^{\tau \tau^{\prime} *}_{\alpha\beta}$ is complex conjugate of $s^{\tau \tau^{\prime}}_{\alpha\beta}$.
After taking integration over $\phi$, the total spin-up/down pumped current in a K-valley becomes: 
\begin{equation}\label{pump2}
  I^{\tau}_{K}=\int_{-\pi/2}^{\pi/2}I^{\tau}_{K}(\phi)\cos(\phi)d\phi.
 \end{equation}
Similarly, For $K'$ valley we get pumped current by replacing $t\rightarrow (-t)$ and $k_y\rightarrow(-k_y)$ in the 
Hamiltonian, Eq.~(1) and wavevectors Eqs.~(9,10) below. The pumped currents in each valley of course depend on the incident angle as well as energy of the  electron. \\
%
{\em Effect of finite temperature:}\\
So far we confined our discussion at zero temperature.
The effects of any non-zero temperature could be easily absorbed by multiplying a
factor $[-df(E)/dE]$ with $I^{\tau}_{K}(\phi)$ and integrating over electron energy\cite{buttiker} as-
\begin{equation}\label{temp}
 I^{\tau}_{K}(\phi)=\int_{0}^{\infty} \big[-\frac{df(E)}{dE}\big]I^{\tau}_{K}(\phi)dE , 
 \end{equation}
 where $\tau=\ua,\da,$  and $f(E)$ is the Fermi-Dirac distribution function. 
 
 \subsection{ The scattering problem: Wave functions and boundary conditions}
Let us consider the case of a spin-up electron with energy $E$, scattered from magnetic impurity at an
incidence angle of $\phi$. The electron can be reflected or transmitted to spin-up/down electron.
We shall start with the inclusion of disorder in the system. We have modeled the system in such way that
the two independent system parameters (magnetic impurity and electrostatic delta potential)
are at the two ends of the system, where disorder is randomly distributed in the strained region.
We also assume that randomly distributed potentials are loclaized in $x-$direction but extended along $y-$direction, i.e., superlattice type potential.

The strength of random potentials are taken in the range of $100-150$  meV-nm.
To obtain the scattering amplitudes, we shall adopt the transfer matrix approach. Transfer matrix connects the wave function amplitudes between
left and right of the scatterer. The wave function for A-sublattice in each strained region for K-valley can be written as:
\begin{figure}
\includegraphics[width=.45\textwidth,height=55mm]{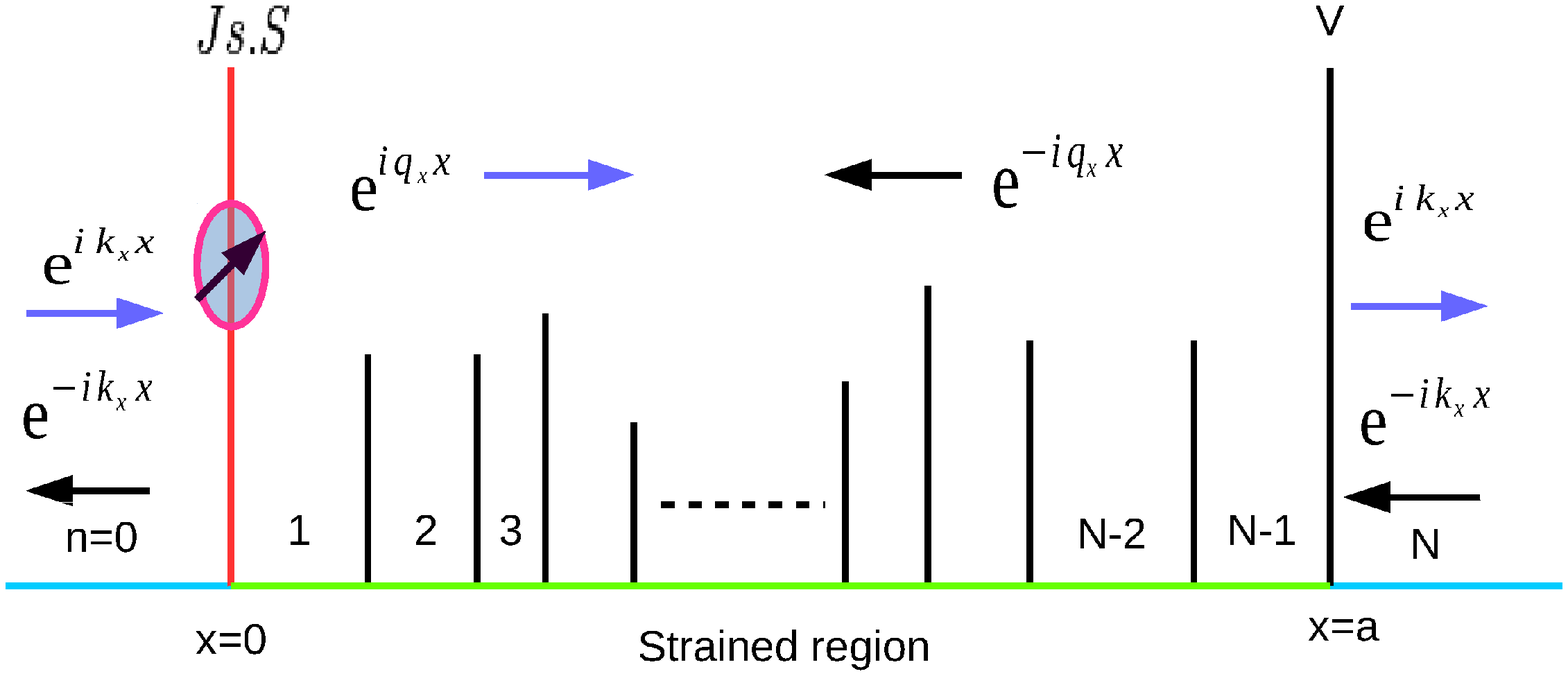}
\caption{Randomly distributed disorder potentials with random strength are confined between two system parameters i.e J and V which are adiabatically modulated.}
\label{Fig9}
\end{figure}
\begin{figure}
\leavevmode
\includegraphics[width=95mm,height=7cm]{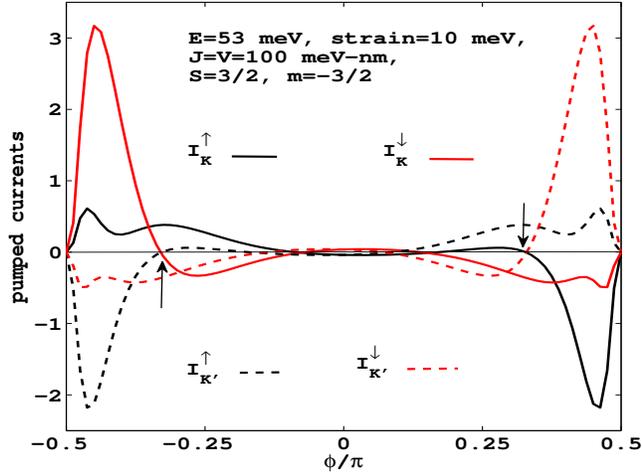}
\caption{Quantum pumped spin up/down currents vs angle of incidence in each valley.
Black arrow signs are used to indicate the quantum spin-valley Hall effect, i.e;
 $I_{K}^{\downarrow}=-I_{K'}^{\uparrow}$ while $I_{K}^{\uparrow}=I_{K'}^{\downarrow}=0$ at a particular angle of incidence, QSVH of first kind.}
 \label{Fig3}
\end{figure}
\begin{figure*}
     \subfigure[Quantum pumped spin up/down currents in each valley Vs. angle of incidence of electron.
 Spin up current ($I_{\ua}$) or spin down current $I_{\da}$)
 in either valley is flowing in opposite direction for a wide range of $\phi$.]
 { \includegraphics[width=0.325\textwidth,height=6.0 cm]{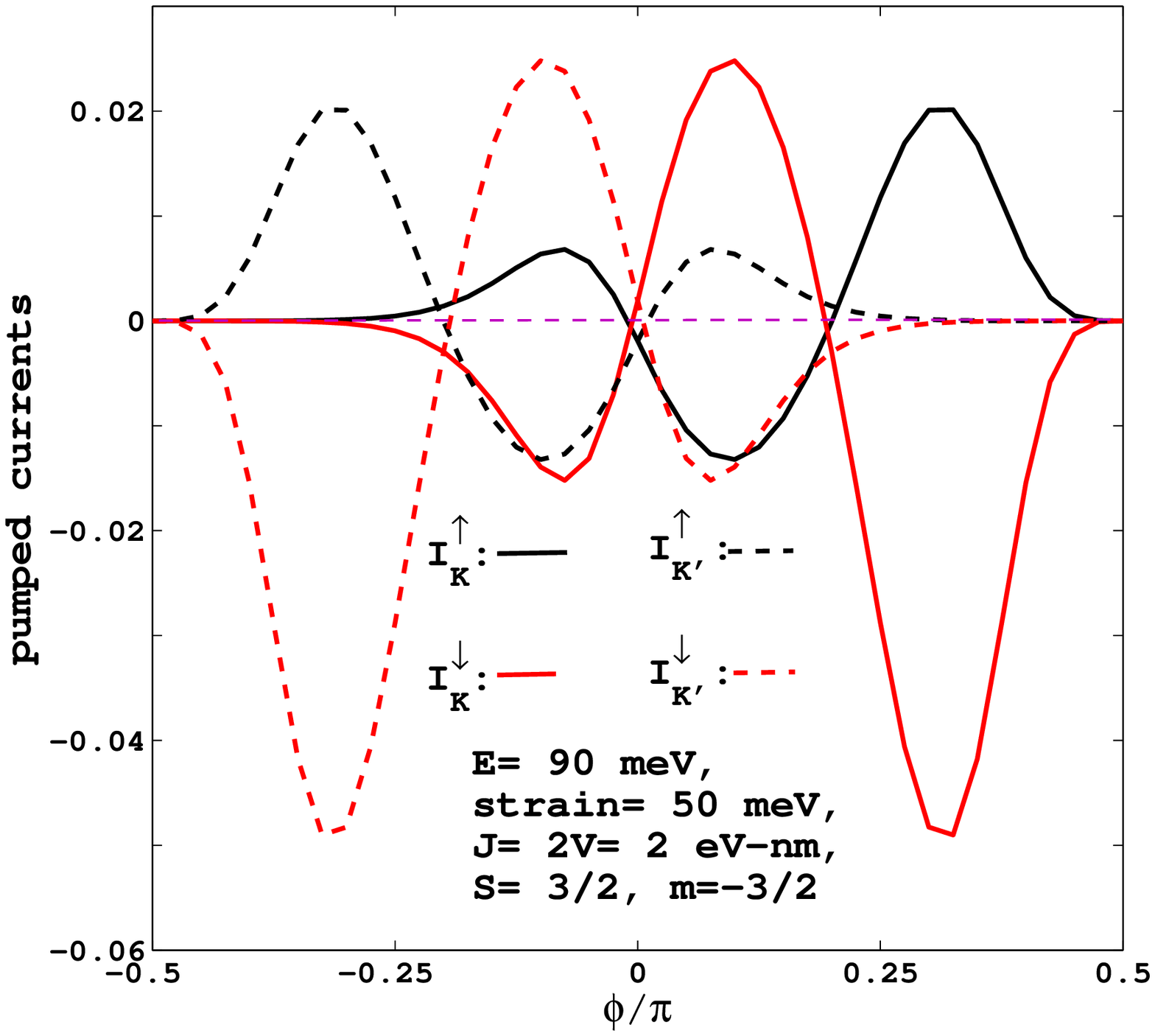}}
   \subfigure[Quantum pumped charge (upper panel) and spin (lower panel) currents in each valley Vs. angle of incidence.
Here, $I_{c/s}=I_{c/s}^{K}+I_{c/s}^{K'}$. The upper panel shows that charge currents in both
 valleys are same in magnitude and flow in exactly opposite direction-QVH, while the lower panel
shows the same for spin currents-QSVH of 2nd kind.]
    {\includegraphics[width=.325\textwidth,height=6.0 cm]{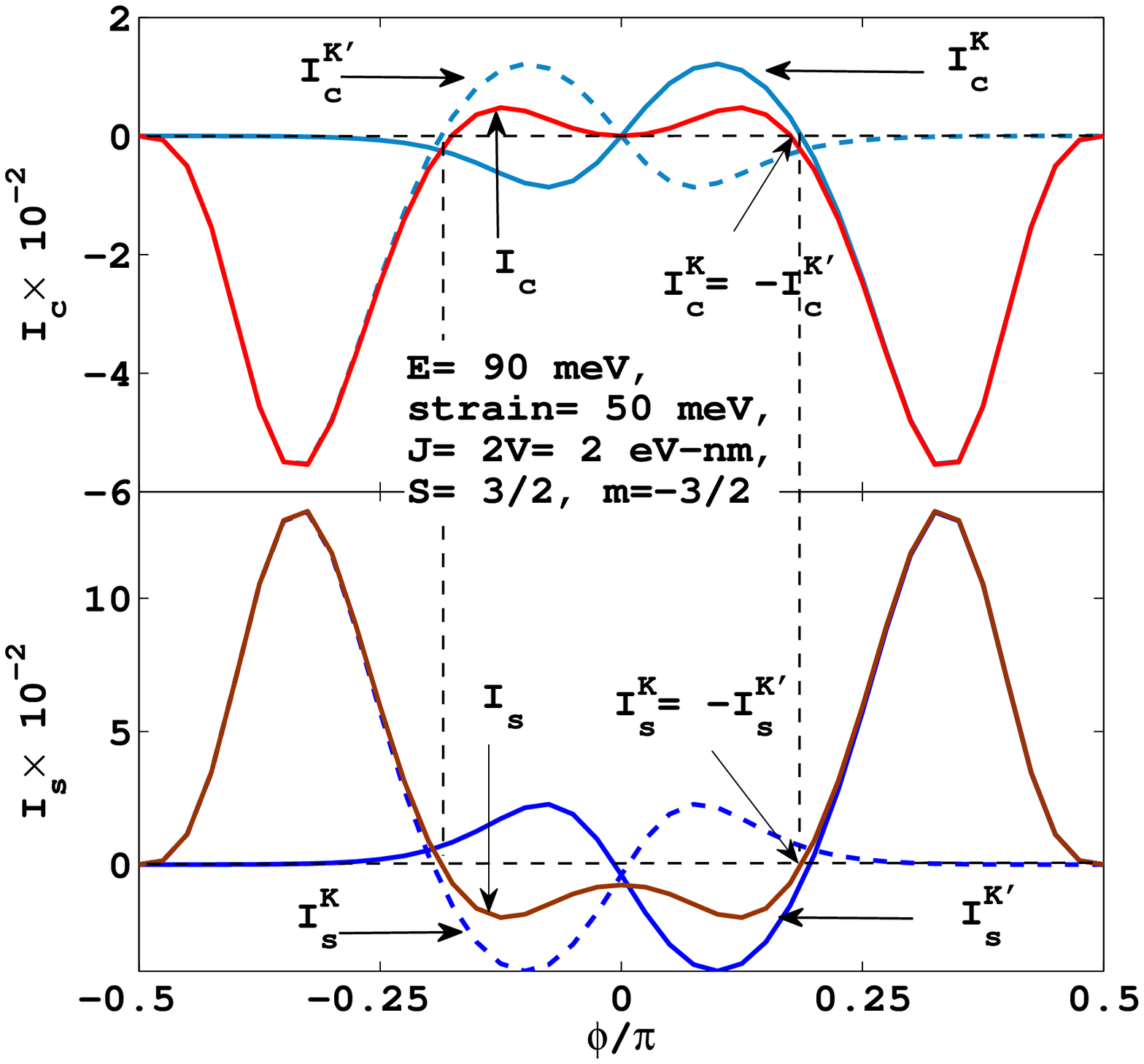}}
    \subfigure[Zoomed portion of (0-$\pi/4$) of Fig. (4b) is shown here. The upper panel shows
the QVH while lower panel shows the QSVH of 2nd kind.]
 { \includegraphics[width=0.325\textwidth,height=6.0 cm]{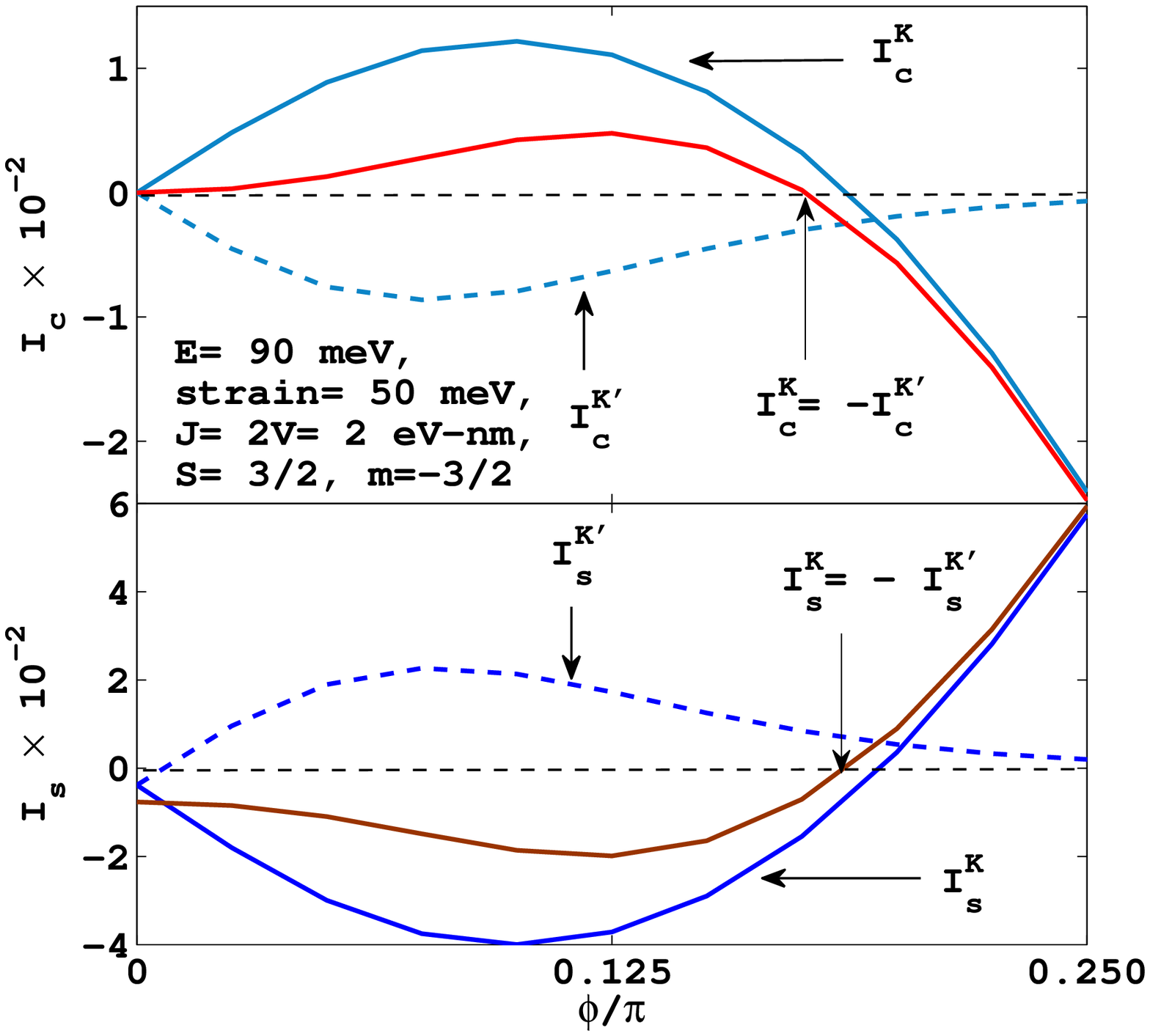}}
  \caption{Pumped spin-valley currents}
  \label{Fig4}
   \end{figure*}
   \begin{figure*}
     \subfigure[Quantum pumped spin up/down currents in each valley Vs. m for $S=3/2$. Parameters are taken as in Fig.\ref{Fig4}
      at an angle of incidence $\phi=\pi/6$ around which  QVH (condition II) and QSVH(2nd kind) (condition IV)  are met when $m=-3/2$.]
 { \includegraphics[width=0.45\textwidth,height=5.3 cm]{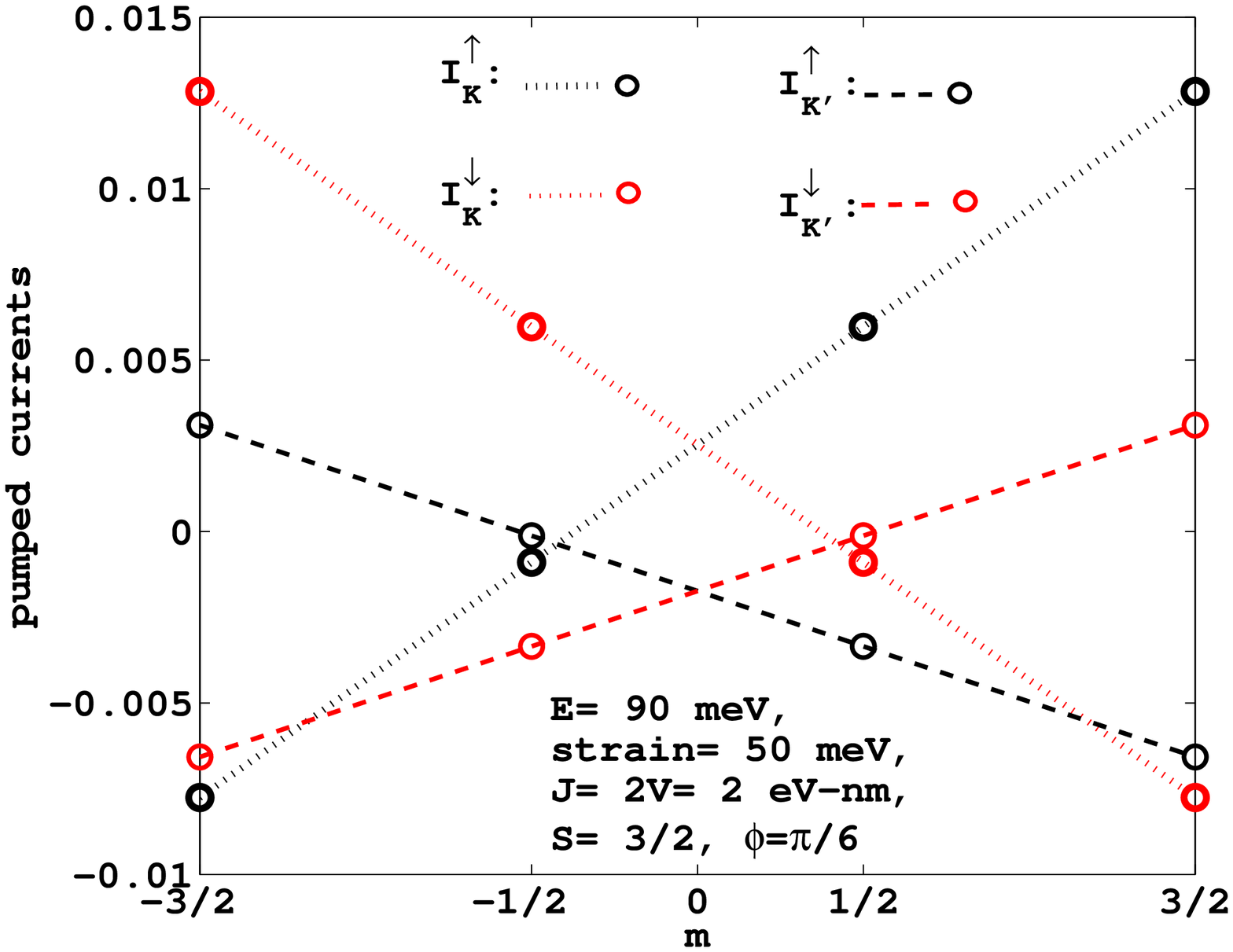}}
   \subfigure[Quantum pumped spin up/down currents Vs m for $S=5/2$. Though same parameters are used as in Fig.\ref{Fig4} except $S=5/2$,
   but $I_{K}^{\uparrow}$ ($I_{K}^{\downarrow}$) and $I_{K'}^{\uparrow}$($I_{K'}^{\downarrow}$) flow in the same direction.]
    {\includegraphics[width=.45\textwidth,height=5.3 cm]{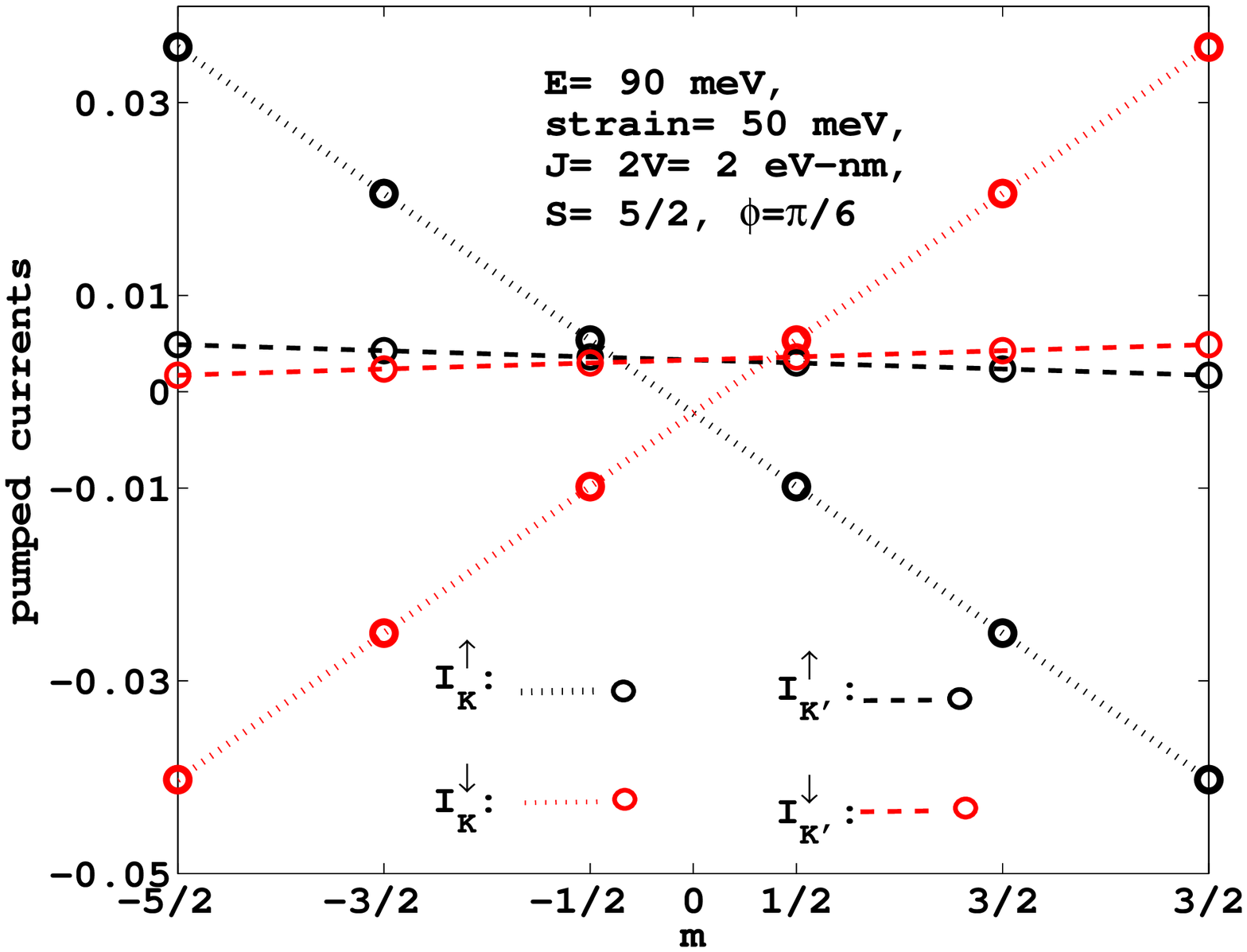}}
  \caption{Quantum pumped spin up/down currents in each valley Vs. magnetic quantum number (m).
 with different S for a particular angle of $\phi$.}
  \label{FigIm}
   \end{figure*}
   
\begin{eqnarray}
 \psi_{n}^{A}(x)&=&
  (A^{\uparrow}_{n}e^{iq_xx}+B_{n}^{\uparrow}e^{-iq_xx})\chi_{\frac{1}{2}}\xi_m\nonumber\\&+&
  (A^{\downarrow}_{n}e^{iq_xx}+B^{\downarrow}_{n}e^{-iq_xx})\chi_{-\frac{1}{2}}\xi_{m+1} .
\end{eqnarray}
and for B-sublattice
\begin{eqnarray}
 \psi_{n}^{B}(x)&=&
  (A^{\uparrow}_{n}e^{iq_xx+i\theta}-B_{n}^{\uparrow}e^{-iq_xx-i\theta})\chi_{\frac{1}{2}}\xi_m\nonumber\\&+&
  (A^{\downarrow}_{n}e^{iq_xx+i\theta}-B^{\downarrow}_{n}e^{-iq_xx-i\theta})\chi_{-\frac{1}{2}}\xi_{m+1}.
\end{eqnarray}
Here $n=1,2,3...(N-1)$ corresponding to different regions bounded by the delta potentials, as shown in Fig. \ref{Fig9}.
The x-component of the momentum vector inside the strained region: $q_x=\sqrt{(E/\hbar v_F)^2-(k_y-t)^2}$.
For the unstrained region i.e; $``n=0"$ and $``n=N"$, $q_x$ has to be replaced by $k_x$, where $k_x=E\cos\phi/(\hbar v_F)$.
The phase factor inside the strained region is defined by $\tan\theta=(k_y-t)/q_x$.
$\xi_m$ is the eigen state of z-component of spin operator of magnetic impurity $S_z$, $ S_z\xi_m=m\xi_m$
with $m$ being the corresponding eigen value. The scattering mechanism is considered as elastic
and the z-component of the total spin remains conserved.
Following the Refs.\cite{menezes,griffiths,bose}, we obtain the 
boundary conditions at the location of two independent time dependent system parameters $J$ and $V$ as: \\
at $x=0$:
\begin{eqnarray}\label{bc1}
 -i\hbar v_{F}[\psi_{1}^{B}(x&=&0)-\psi_{0}^{B}(x=0)]=\frac{J}{2}{\bf s.S}
 [\psi_{1}^{A}(x=0)+\psi_{0}^{A}(x=0)]\nonumber\\
\end{eqnarray}
and
\begin{eqnarray}\label{bc2}
 -i\hbar v_{F}[\psi_{1}^{A}(x&=&0)-\psi_{0}^{A}(x=0)]=\frac{J}{2}{\bf s.S}
 [\psi_{1}^{B}(x=0)+\psi_{0}^{B}(x=0)]\nonumber\\.
\end{eqnarray}
at $x=a$:
\begin{eqnarray}\label{bc3}
 -i\hbar v_{F}[\psi_{N}^{B}(x&=&a)-\psi_{N-1}^{B}(x=a)]=\frac{V}{2} [\psi_{N}^{A}(x=a)+\psi_{N-1}^{A}(x=a)]\nonumber\\
\end{eqnarray}
and
\begin{eqnarray}\label{bc4}
 -i\hbar v_{F}[\psi_{N}^{A}(x&=&a)-\psi_{N-1}^{A}(x=a)]=\frac{V}{2} [\psi_{N}^{B}(x=a)+\psi_{N-1}^{B}(x=a)]\nonumber\\.
\end{eqnarray}
Before proceeding further, we shall mention that spin flipping process is attributed to the interaction term between
the spin of electron (s) and the spin of magnetic impurity (S), ${\bf s.S}=s_zS_z+(1/2)(s^{-}S^{+}+s^{+}S^{-})$as:
$s^{-}S^{+}\left[\begin{array}[c]{c}
       1\\0
       \end{array}\right]
\xi_m=F\left[\begin{array}[c]{c}
         0\\1
        \end{array}\right]
\xi_{m+1}$ and $s^{+}S^{-}\left[\begin{array}[c]{c}
       0\\1
       \end{array}\right]
\xi_m=F'\left[\begin{array}[c]{c}
         1\\0
        \end{array}\right]
\xi_{m-1}$
with $F=\sqrt{(S-m)(S+m+1)}$ and $F'=\sqrt{(S+m)(S-m+1)}$.
Here, $s_{z}$ and $S_{z}$ are the z-components of the spin operator of electron and magnetic
impurity, respectively. $S^{\pm}=S_{x}\pm iS_{y}$ are the raising and lowering operators for magnetic impurity,
and $s^{\pm}=s_{x}\pm is_{y}$ are the same for conduction electron.\\

Following the boundary condition prescribed in Eqs.(\ref{bc1})-(\ref{bc4}), the transfer matrix across the magnetic impurity (at $x=0$)
i.e., between region $``n=0"$ and $``n=1"$ as in Fig. \ref{Fig9} is given as-
\begin{equation}\label{T-matrix}
 \left[\begin{array}[c]{c}A^{\uparrow}_{1}\\A^{\downarrow}_{1}\\B^{\uparrow}_{1}\\B^{\downarrow}_{1}\end{array}\right]=
\mathcal{M}^{[1,0]}\left[\begin{array}[c]{c}A^{\uparrow}_{0}\\A^{\downarrow}_{0}\\B^{\uparrow}_{0}\\B^{\downarrow}_{0}\end{array}\right],
\end{equation}
where $\mathcal{M}^{[1,0]}$, transfer matrix across magnetic impurity, given by $\mathcal{M}^{[1,0]}=\mathcal{M}_{0}^{-1}\mathcal{M}_{1}$ with
\begin{equation}
\mathcal{M}_{0}=
\left[\begin{array}[c]{cccc} \bar{\xi}- iJ^{\prime}m& -iJ^{\prime}F &iJ^{\prime}m-\bar{\xi}_c & -iJ^{\prime}F
                 \\-iJ^{\prime}F &\bar{\xi}+iJ^{\prime}(m+1)& -iJ^{\prime}F &-iJ^{\prime}(m+1)-\bar{\xi}_c\\
            1-iJ^{\prime}m \bar{\xi} & -iJ^{\prime}F\bar{\xi} &1+iJ^{\prime}m \bar{\xi}_c &iJ^{\prime}F\bar{\xi}_c\\-iJ^{\prime}F \bar{\xi}&
            1+iJ^{\prime}(m+1)\bar{\xi}& iJ^{\prime}F\bar{\xi}_c& 1-iJ^{\prime}(m+1)\bar{\xi}_c
           \end{array}\right]\nonumber\\,
\end{equation}
and
\begin{equation} \mathcal{M}_{1}=\left[\begin{array}[c]{cccc} \xi+ iJ^{\prime}m& iJ^{\prime}F &iJ^{\prime}m-\xi_c & iJ^{\prime}F
                 \\iJ^{\prime}F &\xi-iJ^{\prime}(m+1)& iJ^{\prime}F &-iJ^{\prime}(m+1)-\xi_c\\
            1+iJ^{\prime}m \xi & iJ^{\prime}F\xi &1-iJ^{\prime}m \xi_c &-iJ^{\prime}F\xi_c\\iJ^{\prime}F \xi& 1-iJ^{\prime}(m+1)\xi&
                -iJ^{\prime}F\xi_c& 1+iJ^{\prime}(m+1)\xi_c
           \end{array}\right]\nonumber\\
\end{equation}
with $\bar{\xi}=\exp(i\theta)$ and $\bar{\xi}_c=\exp(-i\theta)$, $\xi=\exp(i\phi)$ and $\xi_c=\exp(-i\phi)$.
Also, $J^{\prime}=J/(2\hbar v_F)$. Similarly, the transfer-matrix between ``$n=N$'' and ``$n=N-1$'' at $x=a$, is
\begin{equation}
 \left[\begin{array}[c]{c}A^{\uparrow}_{N}\\A^{\downarrow}_{N}\\B^{\uparrow}_{N}\\B^{\da}_{N}\end{array}\right]=
\mathcal{M}^{[N,N-1]}\left[\begin{array}[c]{c}A^{\uparrow}_{N-1}\\A^{\da}_{N-1}\\B^{\uparrow}_{N-1}\\B^{\da}_{N-1}\end{array}\right],
\end{equation}
where $\mathcal{M}^{[N,N-1]}$ is the transfer matrix across any disorder potential,
expressed as $\mathcal{M}^{[N,N-1]}=\mathcal{M}_{N-1}^{-1}\mathcal{M}_{N}$ with
\begin{equation}
 \mathcal{M}_{N-1}=\left[\begin{array}[c]{cccc}\xi- iV^{\prime}& 0 &iV^{\prime}-\xi_c & 0 \\0 &\xi-iV^{\prime}& 0 &iV^{\prime}-\xi_c\\
            1-iV^{\prime} \xi & 0&1+iV^{\prime} \xi_c &0\\0& 1-iV^{\prime}\xi& 0& 1+iV^{\prime}\xi_c
           \end{array}\right]\nonumber\\,
   \end{equation}
   and
\begin{equation}
 \mathcal{M}_{N}=\left[\begin{array}[c]{cccc} \bar{\xi}+ iV^{\prime}& 0 &iV^{\prime}-\bar{\xi}_c & 0 \\0 &\bar{\xi}+iV^{\prime}& 0 &iV^{\prime}-\bar{\xi}_c\\
            1+iV^{\prime} \bar{\xi} & 0 &1-iV^{\prime} \bar{\xi}_c &0\\0& 1+iV^{\prime}\bar{\xi}& 0& 1-iV^{\prime}\bar{\xi}_c
           \end{array}\right].
\end{equation}
Here, $V^{\prime}=V/(2\hbar v_F).$
Since electrostatic potential at x=a ( acting as a system parameter) and disorder potential
are both modeled as delta function potential, the transfer- matrix for any arbitrary interface between $x=0 $
and $x=a$ has also the same matrix elements as $\mathcal{M}^{[N,N-1]}$.
After some straight forward algebraic manipulation, we construct the total transfer-matrix which connect s
the wave function amplitudes of extreme left and right as\cite{griffiths}
\begin{equation}
 \left[\begin{array}[c]{c}A^{\uparrow}_{N}\\A^{\da}_{N}\\B^{\uparrow}_{N}\\B^{\da}_{N}\end{array}\right]=
\mathcal{M}\left[\begin{array}[c]{c}A^{\uparrow}_{0}\\A^{\da}_{0}\\B^{\uparrow}_{0}\\B^{\da}_{0}\end{array}\right],\nonumber\\
\end{equation}
where
\begin{equation}
       \mathcal{M}=\mathcal{M}^{[N,N-1]}\mathcal{M}_{free}^{[N-1]}\mathcal{M}^{[N-1,N-2]}\mathcal{M}_{free}^{[N-2]}
       .....\mathcal{M}_{free}^{[1]}\mathcal{M}^{[1,0]}
\end{equation}
with $\mathcal{M}^{n}_{free}$ being the propagation matrix between any two successive disorder potential, which is given by
\begin{equation}\label{free}
 \mathcal{M}^{n}_{free}=\left[\begin{array}[c]{cccc}  e^{iq_xd_n}& 0& 0 &0 \\0 &e^{iq_xd_n} & 0 &0\\
            0 & 0 &e^{-iq_xd_n} &0\\0& 0& 0&e^{-iq_xd_n}
           \end{array}\right]
\end{equation}
with $d_n$ is the spatial gap between two successive disorder potentials. Also,
$\mathcal{M}^{[N,N-1]}$ is the transfer-matrix which connect the wave function amplitudes between the regions ``N-1" and ``N".
To calculate the reflection and transmission amplitudes, we shall use the relation between scattering matrix
and transfer-matrix as \cite{griffiths}
\begin{eqnarray}
 S&=&\frac{1}{\mathcal{M}_{22}}\left[\begin{array}[c]{cc}\mathcal{M}_{21}& \mathcal{I} \\ \mathcal{I}\det\mathcal{M} &\mathcal{M}_{12}
                       \end{array}\right],
  \end{eqnarray}
  with
  \begin{eqnarray}
   \mathcal{M}=\left[\begin{array}[c]{cc}  \mathcal{M}_{11}& \mathcal{M}_{12} \\ \mathcal{M}_{21} & \mathcal{M}_{22}
                       \end{array}\right]=\left[\begin{array}[c]{cccc}  m_{11}& m_{12}&m_{13}&m_{14}\\ m_{21}& m_{22}&m_{23}&m_{24} \\ m_{31}& m_{32}&m_{33}&m_{34} \\ m_{41}& m_{42}&m_{43}&m_{44}
                       \end{array}\right],
\end{eqnarray}
as is obvious from Eq.~(25), $\mathcal{I}, \mathcal{M}_{11}, \mathcal{M}_{12}, \mathcal{M}_{21}, \mathcal{M}_{22} $ are all $2\times2$ matrices.
The reflection amplitude (to the left, as we are calculating pumping current in the left lead)
\begin{equation}
 r=-\frac{\mathcal{M}_{21}}{\mathcal{M}_{22}}=\left[\begin{array}[c]{cc}  r_{\uparrow\uparrow}& r_{\uparrow\downarrow} \\ 
 r_{\downarrow\uparrow} &r _{\downarrow\downarrow}
                       \end{array}\right]
                       \end{equation}
                      
and transmission amplitude from right to left is 
\begin{equation}
 t=\frac{1}{\mathcal{M}_{22}}=\left[\begin{array}[c]{cc}  t_{\uparrow\uparrow}& t_{\uparrow\downarrow} \\ 
 t_{\downarrow\uparrow} &t _{\downarrow\downarrow}
                       \end{array}\right].
\end{equation}
The scattering amplitudes obtained by the above method can be directly used in Eq. (\ref{pump}) to obtain the  quantum pumping current. Now we can
recover the situation of disorder free pumped current by
using transfer matrix as $\mathcal{M}=\mathcal{M}^{[2,1]}\mathcal{M}_{free}^{1}\mathcal{M}^{[1,0]}$,
where $\mathcal{M}^{[2,1]}$ and $\mathcal{M}^{[1,0]}$ would become the transfer-matrix across 
the electrostatic potential and magnetic impurity, respectively. And, $d_1$ would become $a$ in Eq. (\ref{free}).
By solving this scattering problem numerically, 
we obtain different scattering amplitudes which obey probability conservation
$\mid t_{\ua\ua}\mid^2+\mid r_{\ua\ua}\mid^2+\mid t_{\da\ua}\mid^2+\mid r_{\da\ua}\mid^2=1$
for a particular angle of incidence $\phi$ and for particular spin (here, $\ua$) incident.
In experiment, graphene electrons can be incident at a particular angle by means
of beam collimation techniques (discussed in conclusion also). 
To focus graphene electron at particular angle without any spatial spreading, periodic potential
can be used suitably as proposed by Park, et. al., in Ref.\cite{park}.

Similarly for the case of spin-down incident electron from the left side, we can get scattering amplitudes.
This procedure can be repeated appropriately for spin-up/down electron coming from right side. 
We repeat this for for $K^{\prime}$-valley by chosing $t\rightarrow (-t)$ and $k_y\rightarrow (-k_y)$
in the Hamiltonian and corresponding wavefunctions.

 \section{Results and discussion} 
 To calculate pumped current for different spin components, we use the formula- Eq. (6) of Theory section.
 First, we have plotted different components of spin pumped currents (in units of $I_0$) i.e;
 spin-up ($I_\ua$), spin-down ($I_\da$), spin current ($I_s=I_\ua-I_\da$) and charge
 current $(I_c=I_\ua+I_\da$), in K and K' valley, shown in Fig. \ref{Fig3}.
 
\begin{figure*}
     \subfigure[Pumped currents in each valley after integration over angle from $-\pi/2$
 to $\pi/2$ (here, $I_K=I_{K'}$ because of the time reversal symmetry).
 There is pure spin current, satisfying the condition (Ia).]
 { \includegraphics[width=0.45\textwidth,height=5 cm]{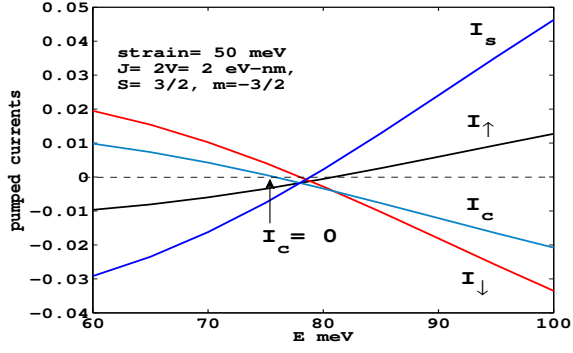}}
   \subfigure[Pumped currents Vs m for $S=3/2$ after integrating over angle of incidence. We
   use parameters as in Fig(\ref{Fig6}a) at $E=74.6$ meV  where Condition (Ia) is satisfied.]
    {\includegraphics[width=.45\textwidth,height=5 cm]{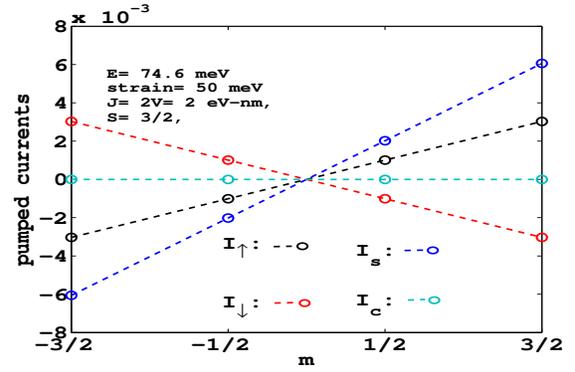}}
  \caption{Quantum pumped currents in each valley Vs. magnetic quantum number (m).}
  \label{Fig6}
   \end{figure*}

\begin{figure}
\leavevmode
\includegraphics[width=95mm,height=7cm]{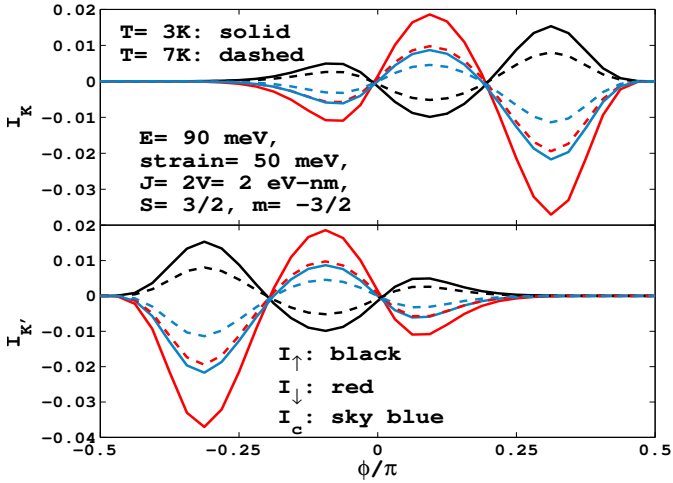}
\caption{Effect of temperature on pumped current in each valley Vs angle of incidence.}
\label{Fig7}
\end{figure}

\begin{figure}
\leavevmode
\includegraphics[width=90mm, height=7cm]{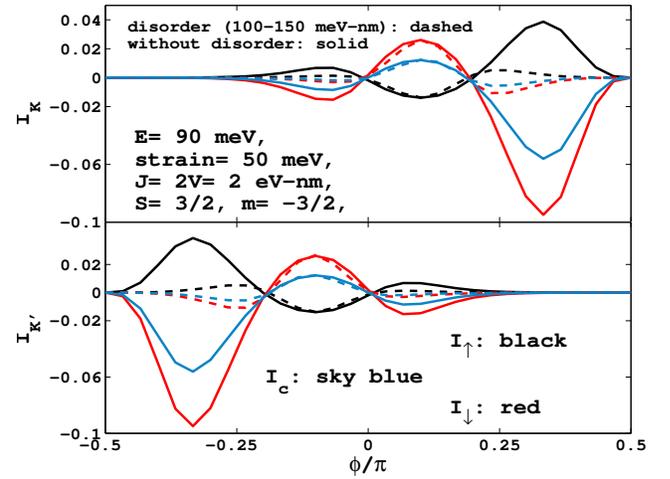}
\caption{Effect of disorder on pumped currents Vs angle of incidence.}
\label{Fig8}
\end{figure}

 We have chosen parameters: the spatial separation between  the magnetic impurity and the electrostatic delta potential $a= 40$ nm,
 spin of the molecular magnet $S=3/2$, and $m = -3/2$ in all figures~\ref{Fig3}-\ref{Fig8}. Here, $m$
 is the eigen value of $S_z$, the z-component of the spin operator of magnetic impurity.
 Other parameters are mentioned in the figures.
 
 In Fig.~\ref{Fig3}, we see that at a particular angle of incidence, one valley carries spin up
 current while other valley carries spin down current with same magnitude but in opposite direction.
 We notice that around $\phi=\pi/3$, $I_{K'}^{\downarrow}=I_{K}^{\uparrow}=0$ but $I_{K'}^{\uparrow}=-I_{K}^{\downarrow}$, 
 satisfying condition (IIIa)-QSVH of 1st kind. Similarly around $\phi=-\pi/3$, we see
 $I_{K'}^{\uparrow}=I_{K}^{\downarrow}=0$ but $I_{K}^{\uparrow}=-I_{K'}^{\downarrow}$,
 satisfying the condition (IIIb)-QSVH of 1st kind.
  
%
%

 The spin current $I_{s}$ and charge current $I_{c}$ corresponding to Fig.~\ref{Fig4}(a) are shown in Fig.~\ref{Fig4}(b).
 In the upper panel of Fig.~\ref{Fig4}(b), we find that charge  current in K and K'-valley are same in magnitude
 but opposite in direction, satisfying the condition (II) i.e; quantum valley Hall effect (QVH).
 The lower panel of Fig.~\ref{Fig4}(b) shows that spin current  in K and K'-valley are same but opposite
 in direction, satisfying the condition (IV) i.e; quantum spin-valley Hall effect (QSVH) of 2nd kind.
 A zoomed portion (0-$\pi/4$) of Fig.~\ref{Fig4}(b) is given in Fig.~\ref{Fig4}(c). For $S=3/2$, there are four possible values of $m$ $(-3/2,-1/2,1/2,3/2)$. In Fig.\ref{FigIm}(a), we show how
different components of quantum pumped current varies with $m$ for a particular angle $\phi=\pi/6$. Similar plot
is also given for $S=5/2$ in Fig.~\ref{FigIm}(b), which indicates that pumped currents are very sensitive to $m$.
 
 We also consider the case of all angle incidence, i.e.,
 we integrate over the angle of incidence $\phi$ in Eq.(\ref{pump2}),
 we plot pumped currents versus energy in Fig.~\ref{Fig6}(a). Because of the time reversal symmetry, pumped currents in
 both valleys would be identical. Here, we see that pure spin current appears at a certain energy 
 satisfying the condition (Ia). In the Fig.~\ref{Fig6}(b), it is shown that we get a pure spin current 
 regardless of the $m$ value, satisying condition 1a of  box.\\
 
{\em  \underline {Effect of temperature and disorder on QSVH:}}\\
Temperature has a very significant influence on transport properties, especially 
on the magnitude of transport coefficients. Here, we look at how temperature can affect the
pumped spin valley currents. We plot pumped currents versus the angle of incidence for two 
different non-zero temperatures in Fig.~\ref{Fig7}, for which we used  Eq. (\ref{temp}).
It  is found that pumped currents are damped with increase in temperature, however the 
location of QVH or QSVH remains intact, temperature cannot shift the incidence angle  ehere QVH or QSVH occur.

The study of disorder effects on transport properties has been always important, as 
disorder is always present in the electronic system. Here, we intend to examine
how pumped currents get affected by random potential. For this, we treat the random potentials as delta like potential and 
solved the scattering problem by transfer matrix approach and then use Eq.~(\ref{pump}) to calculate the pumped currents.
The presence of randomly distributed impurities/adatoms/vacancies modeled by 
the delta potentials in the system can suppress the pumped currents which is shown in Fig.~\ref{Fig8}.
We find that magnitude of pumped currents are damped due to the presence of randomness, but again no change in
the location where QSVH appears. However, very strong disorder may lead to the non-trivial changes.

A tabular representation of our findings is given below:

\begin{table}[h]
\centering
\colorbox{white}{
\begin{tabular}{|c |c |c |c|c|}
\hline\hline
Figure($\downarrow$) & QVH & pure spin & QSVH  & QSVH \\
&  &current in & of 1st kind & of 2nd kind\\
& & each valley & &\\
(Condition$\rightarrow$)&(II) &(Ib/Ia)& (IIIa/IIIb)& (IV)\\
\hline
 \ref{Fig3} & ------  & ------- & present &------\\
&  &  & &\\
\hline
 \ref{Fig4}b & present & present (Ib) & absent& present\\
 &  & & &\\
 \hline
 \ref{Fig6} & ------ & present (Ia)& ------& ------ \\
     & &  & & \\
\hline
 \ref{Fig7} and \ref{Fig8} & same as \ref{Fig4}b & same as \ref{Fig4}b & same as \ref{Fig4}b& same as \ref{Fig4}b\\
         & but damped & but damped & but damped& but damped\\
\hline
\end{tabular}
}
\caption{Summary of the results}
\end{table}
\section{Experimental realization and Conclusions}
The 1D electrostatic delta potential can be realized by placing a series of several adatoms which can be
adiabatically modulated by a gate voltage. This potential is acting as a system parameter only,
it has nothing to do with spin/valley degree of freedom,
so one can also use a thin rectangular potential barrier instead of delta potential.
The 1D chain of magnetic impurity with $S=3/2$ can be used as the other system parameter,  
experimental feasibility of these kind of wires is already established\cite{wire1,wire2}. The strength of 
exchange interaction can be varied by tuning the magnetic field of a ferromagnet placed on
top of the magnetic chain.
As we have shown that QSVH or QVH is observable at a particular angle of electron incidence,
focusing of electron at particular angle is very important which can be realized by means of beam collimation\cite{park}.
The propagation of graphene electron beam  without any spatial spreading or diffraction can be
experimentally realized by applying 1D spatial periodic potential,
here no external magnetic or electric field is required. This is called super beam collimator\cite{park}.
The phenomena of super beam collimation is described as follows- Under the influence of 1D periodic potential,
group velocity of low energy graphene carriers becomes anisotropic.
By suitably controlling the 1D potential, the extreme anisotropy in velocity can be realized
giving us  electrons on demand at a particular angle of incidence.\\

In conclusion, we have proposed a graphene based device to observe quantum spin valley Hall effect by
adiabatically modulating a magnetic impurity and an electrostatic potential embedded in a monolayer of strained graphene.
In the same device, we have also shown the appearance of quantum valley Hall effect, pure spin 
current generation in each valley and quantum valley Hall effect with pure spin current (QSVH of 2nd kind).
We also examined the effects of temperature and disorder on pumped currents. In future, this work will be
extended further to study the spin-valley dependent electro-optic like
phenomena in strained graphene.

\section*{Acknowledgements}
This work was supported by funds from Dept. of Science and Technology (Nanomission), Govt. of India, Grant No. SR/NM/NS-
1101/2011. Authors thank Arjun Mani, School of Physical Sciences, NISER, Bhubaneswar for useful discussions.
\section{Competing financial interests statement} Authors have no competing financial interests to disclose.

\end{document}